\documentstyle[preprint,aps]{revtex}
\begin{document}
\draft
\preprint{\vbox{\hbox{SNUTP-94/65}\hbox{KWUTP-94/4}}}

\title{Bogomol'nyi equations for solitons in Maxwell-Chern-Simons\\
gauge theories with the magnetic moment interaction term}

\author{Taejin Lee}

\address{
Department of Physics, Kangwon National University, Chuncheon 200-701, KOREA }

\author{Hyunsoo Min}

\address{
Department of Physics, Seoul City University, Seoul 130-743, Korea}

\maketitle
\begin{abstract}
Without assuming rotational invariance, we derive Bogomol'nyi equations for
the solitons in the abelian Chern-Simons gauge theories with the anomalous
magnetic moment interaction. We also evaluate the number of zero modes
around a static soliton configuration.
\end{abstract}


\narrowtext


In recent years there have been numerous studies on the
soliton solutions of Chern-Simons gauge theories since the initial finding that
they
carry electric charge \cite{cssol} as well as magnetic flux in contrast to the
old
electrically neutral Nielsen-Olesen vortices \cite{novor}. To study properties
of solitons, it is often useful to consider the special
cases of the Bogomol'nyi limit first. In these cases, the original second order
field equations reduce to the first order ones which are usually called
Bogomol'nyi equations \cite{lg} or self-duality equations. Then the solitons
become noninteracting and static multi-soliton solutions may be expected.
These Bogomol'nyi cases are realized with a very special choice of parameters
and
a specific form of the scalar potential: See ref.\cite{lg}, ref.\cite{csvor},
and ref.\cite{mcsvor} for the Bogomol'nyi equations of the
Landau-Ginzburg model, the Chern-Simons gauge theory, and the
Maxwell-Chern-Simons theory respectively.

Very recently, Torres \cite{torr} considered soliton solutions in a
Maxwell-Chern-Simons gauge theory with an additional magnetic moment
interaction term. If one wishes to consider a low energy effective theory
containing at most second order derivative terms, one may not exclude
such a magnetic moment interaction. This magnetic moment
interaction term in Chern-Simons gauge theory has been considered
also in some other context \cite{other}. Assuming the rotational invariance,
Torres studied the lower bound of the energy functional. And he
found a simple second order potential for the Bogomol'nyi limit.
Shortly after, Ghosh \cite{gosh} extended Torres' work to a more generalized
Maxwell-Chern-Simons theories. However, since their analyses strongly rely upon
the assumption of the rotational invariance, it is not clear whether
solutions to their `radial' Bogomol'nyi equations correspond to the lowest
energy configurations. For instance, by considering angular variations,
one may be able to reduce the energy further.

In the present work we will derive the Bogomol'nyi equation
{\it without assuming the rotational invariance}, thus putting the works
of Torres and Ghosh on a firmer base. We also evaluate the number of parameters
contained in general solutions of the
Bogomol'nyi equations by studying the fluctuations around soliton
configurations.

Let us start with the model of Torres. It is based on the Lagrangian density
\begin{equation}
{\cal L}=-{\frac{1 }{4}} F_{\mu\nu}F^{\mu\nu}
 +{\frac{\kappa}{2}}\epsilon^{\mu\nu\alpha}
 A_\mu\partial_\nu A_\alpha -\left|{\cal D}_\mu
\phi\right|^2 - \kappa^2|\phi|^2\label{model}
\end{equation}
with ${\cal D}_\mu \phi=\left[\partial_\mu-ieA_\mu  -ig f_\mu
\right] \phi$, where $f_\mu \equiv {1 \over 2}\epsilon_{\mu\nu\alpha}
F^{\nu\alpha}$ is the dual field strength.
Note that the potential has a very specific form, viz., $V=\kappa^2|\phi|^2$.
The equations of motion for the above Lagrangian density are
\begin{mathletters}
\label{eqm:all}
\begin{eqnarray}
{\cal D}_\mu  {\cal D}^\mu \phi &=& -{\partial V \over \partial \phi^*}, \\
\epsilon_{\mu\nu\alpha} \partial_\nu \label{eqm:a}
(f^\alpha+g  J^\alpha) &=& -\kappa(f^\mu+{e \over\kappa}J^\mu) \label{eqm:b}
\end{eqnarray}
\end{mathletters}
with $ J_\mu=-i(\phi^*{\cal D}_\mu\phi-{\cal D}_\mu \phi^*\phi)$.
As Torres observed, when the relation
\begin{equation}
g={\frac{e}{\kappa}},
\end{equation}
holds, any solution to the first order equation
\begin{equation}
f^\mu+g J^\mu=0 \label{first}
\end{equation}
satisfies automatically the second order field equation (\ref{eqm:b}).

In order to derive the Bogomol'nyi equations we consider the energy
functional
\begin{eqnarray}
E=\int d^2 x & & \Biggl[
{1 \over 2} (1-2g^2|\phi|^2)(E_i^2+B^2) \nonumber \\
& &+ |D_0\phi|^2 + |D_i\phi|^2+\kappa^2|\phi|^2
\phantom{1 \over 2}\Biggr]. \label{energy}
\end{eqnarray}
Here it is assumed that $1 > 2g^2|\phi|^2$ to ensure the positivity.
Defining a gauge invariant potential
\begin{equation}
\bar {A}^\mu=A^\mu-{\frac{1}{e}}\partial ^\mu {\rm Arg}(\phi),
\end{equation}
we may rewrite Eq.(\ref{first}) as
\begin{mathletters}
\label{re:all}
\begin{eqnarray}
(1-2g^2|\phi|^2)B =2ge|\phi|^2\bar A^0, \label{re:a} \\
(1-2g^2|\phi|^2)E_i=-\epsilon_{i j}2ge|\phi|^2\bar A_j.\label{re:b}
\end{eqnarray}
\end{mathletters}
Using these equations, we may cast
the part of the energy functional involving the electric field and
the spatial derivative of $\phi$ only into the following form:
\begin{eqnarray}
E_A&=&\int d^2 x\left[{1\over 2}(1-2g^2|\phi|^2)E_i^2
+|D_i\phi|^2\right] \nonumber\\
&=&\int d^2x\left[ (\partial_i|\phi|)^2
+{ e^2(\bar {A}_i)^2|\phi|^2\over 1-2g^2|\phi|^2}\right].
\end{eqnarray}

Now we make use of the relation
\begin{equation}
E_A = \int d^2x\left|\partial_\pm|\phi|
-i{e\bar A_\pm\over\sqrt{1-2g^2|\phi|^2}}|\phi|\right|^2
\pm\int d^2 x {e\over\sqrt{1-2g^2|\phi|^2}}
\epsilon_{i j}\bar A_i\partial_j|\phi|^2\label{obs}
\end{equation}
where $\partial_\pm=\partial_1\pm i\partial_2$
and $\bar A_\pm=\bar A_1 \pm i\bar A_2$.
Using Eq.(\ref{obs}) and integration by parts brings us the expression
\begin{mathletters}
\label{bound:all}
\begin{eqnarray}
E &=& \int d^2x \Biggl[\Biggl|\partial_\pm|\phi|
-i{e \over \sqrt{1-2g^2|\phi|^2}}
\bar A_\pm|\phi|\Biggr|^2  +(\partial_t|\phi|)^2 \nonumber \\
 & & +{1-2g^2|\phi|^2\over 4g^2|\phi|^2}
(B\mp {2e|\phi|^2 \over \sqrt{1-2g^2|\phi|^2}})^2 \Biggr]
\pm {e \over g^2}\Phi \label{bound:a} \\
&\ge& |{e \over g^2}\Phi|, \label{bound:b}
\end{eqnarray}
\end{mathletters}
where $\Phi\equiv \int d^2x B$ denotes the magnetic flux.
Therefore, the energy functional
is bounded from below and the inequality in Eq.(\ref{bound:b}) is saturated
by a static configuration satisfying the following Bogomol'nyi equations
\begin{mathletters}
\label{bogo:all}
\begin{eqnarray}
B\mp {2e|\phi|^2\over \sqrt{1-2g^2|\phi|^2}}=0, \label{bogo:a} \\
\partial_\pm|\phi|-i{e\over \sqrt{1-2g^2|\phi|^2}}
\bar A_\pm|\phi|=0.\label{bogo:b}
\end{eqnarray}
\end{mathletters}
Eq.(\ref{bogo:b}) implies that $\bar A_i$ can be determined
in terms of matter field as
\begin{equation}
e\bar A_i=\pm\epsilon_{i j} \sqrt{1-2g^2|\phi|^2}\partial_j \ln
|\phi|. \label{deter}
\end{equation}
Then it follows from Eq.(\ref{bogo:a}) and Eq.(\ref{deter}) that
\begin{equation}
\nabla^2\ln|\phi|={\frac{2 }{1- 2g^2|\phi|^2}} \left[-e^2|\phi|^2+g^2(%
\partial_i|\phi|)^2\right].
\end{equation}
If we assume the rotationally invariant form for $\phi$, this
equation reduces to the corresponding equation of Torres \cite{torr}.
One may refer to his numerical analysis of the corresponding equation.
The solution describes a nontopological soliton characterized by the integer
$n$
and a real valued constant $\alpha$. Here $n$ and $\alpha$ are determined by
the behaviors of $|\phi|$ near and far from the origin respectively, i.e.,
$|\phi|\rightarrow r^n$ as $r\rightarrow 0$ and
$|\phi|\rightarrow r^{-\alpha}$ as $r\rightarrow \infty$.
The soliton carries the magnetic flux $\Phi=(2\pi/e)(n+\alpha)$ as well as
the electric charge $Q=\int d^2x J^0 $. They are  related to each other through
the  relation $\kappa\Phi=-eQ$ which is the spatial
integration of the time component of Eq.(\ref{first}), the `Gauss law'.
Noting that the right hand side of the bound in Eq.(\ref{bound:b})
can be written as $|{\kappa \over e} Q|$, one can easily see that
the nontopological soliton is  marginally stable against radiation of
the charged elementary excitations of mass $|\kappa|$ and charge $e$.

How many free parameters enter the general solution
of the Bogomol'nyi equations (\ref{bogo:all})?
To answer this question, we may consider fluctuations around the solution
discussed above. Varying Eq.(\ref{bogo:a}), one finds that
the vector fluctuations are fixed by the scalar fluctuation
\begin{equation}
e\delta\bar A_i=\pm\epsilon_{ij}\sqrt{1-2g^2|\phi|^2}\partial_j
{\delta\phi \over |\phi|}.
\end{equation}
The scalar fluctuation $\delta|\phi|$ is determined by the
equation
\begin{eqnarray}
\left[\nabla^2+{4e^2|\phi|^2 \over (1-2g^2|\phi|^2)}
\left(1-{g^2 \over e^2}(\partial_i \ln |\phi|)^2\right) \right]
{\delta|\phi| \over |\phi|} \nonumber \\
-{4e^2|\phi|^2 \over (1-2g^2|\phi|^2)}
(\partial_i \ln |\phi|)\left(\partial_i {\delta|\phi| \over |\phi|}\right)
=0.
\end{eqnarray}
The forms of these fluctuation equations near and far from the origin
are quite analogous to those of the Chern-Simons soliton case considered in
ref.\cite{jlw}. Based on a similar analysis, we may conclude
that the number of free parameters in
the general solution is $2(n+[\alpha])$ where $[\alpha]$ is
the greatest integer less than $\alpha$.

We now turn to the Ghosh's generalized Maxwel-Chern-Simons
theory \cite{gosh}, which reduces to the model considered
by Lee and Nam \cite{lm} if the magnetic moment interaction term is switched
off.
It is described by the following Lagrangian density
\begin{equation}
{\cal L}=-{\frac{1 }{4}}G(|\phi|^2) F^2_{\mu\nu}
 +{\frac{\kappa}{2}}\epsilon^{\mu\nu\alpha}
 A_\mu\partial_\nu A_\alpha
 -\left| {\cal D}_\mu  \phi \right|^2 - V ,
\end{equation}
where the covariant derivative is
\begin{equation}
{\cal D}_\mu \phi=(\partial_\mu-ieA_\mu  -ig G(|\phi|^2)f_\mu)\phi.
\end{equation}
The explicit form of the potential $V$ for the Bogomol'nyi limit
will be given later.

The field equations for the vector field $A_\mu$ can be written as
\begin{equation}
\epsilon^{\mu\nu\alpha}
\partial_\nu
G(f_\alpha+g J_\alpha)=
-\kappa(f^\mu+{e \over\kappa}J^\mu).
\end{equation}
So, as in the previous case (see Eq.(\ref{model})),
we can obtain a solution to this second order differential equation
by solving the first order  equation
\begin{equation}
f^\mu+g J^\mu=0 \label{feq}
\end{equation}
if the coupling constant $g$ has the value $g=e/\kappa$.
We may rewrite this equation (\ref{feq}) as
\begin{mathletters}
\label{rfeq:all}
\begin{eqnarray}
(1-2g^2|\phi|^2 G)B  &=&2g e|\phi|^2\bar A^0, \label{rfeq:a}\\
(1-2g^2|\phi|^2 G)E_i &=&-\epsilon_{i j}2g e|\phi|^2\bar A_j. \label{rfeq:b}
\end{eqnarray}
\end{mathletters}

The energy functional for this system is given by
\begin{equation}
E=\int d^2 x \left[
{1\over 2}G(1-2g^2|\phi|^2 G)(E_i^2+B^2)
+|D_0\phi|^2 +|D_i\phi|^2 + V \right].
\end{equation}
Then, using Eqs.(\ref{rfeq:a}) and (\ref{rfeq:b}) and
the identity similar to Eq.(\ref{obs}),
we may express this energy functional as
\begin{eqnarray}
E =& & \int d^2x \Biggl[ \Biggl|\partial_\pm|\phi|
-i{e \over \sqrt{1-2g^2|\phi|^2 G}}
\bar A_\pm|\phi|\Biggr|^2
+(\partial_t|\phi|)^2+(e\bar A^0)^2     \nonumber  \\
& &\phantom{ \int d^2x [}
 +{1\over 2}(1-2g^2|\phi|^2G)B^2 +V
 \pm e \epsilon_{i j}\bar A_i\partial_j{\cal F}(|\phi|^2)
\Biggr].\label{efn}
\end{eqnarray}
where ${\cal F}(|\phi|^2)$ is defined by
\[
\partial_i{\cal F}(|\phi|^2)=
(\partial_i |\phi|^2)/\sqrt{1-2g^2|\phi|^2 G}.
\]
Integrating by parts, we rewrite the last term in Eq.(\ref{efn}) as
\begin{equation}
\int d^2x e\epsilon_{i j}\bar A_i\partial_j {\cal F}=
\int d^2x \left[e({\cal F}-{\cal F_\infty})
(B-\epsilon_{i j} \partial_i\partial_j{\rm Arg}(\phi))
\right] \label{last}
\end{equation}
where ${\cal F}_\infty$ is the value of ${\cal F}$ at $r=\infty$.
Note that since ${\rm Arg}(\phi)$ is not a single valued function,
$\epsilon_{i j} \partial_i\partial_j{\rm Arg}(\phi)$ does not vanish
but has the value $2\pi \sum_p \delta({\bf x}-{\bf x}_p)$ where
${\bf x}_p$ denotes the zeros of $\phi$.

The energy functional now takes the following form
\begin{eqnarray}
E&=& \int d^2x  \Biggl[ (\partial_t|\phi|)^2
  +\Biggl|\partial_\pm|\phi|
  -i{e \over \sqrt{1-2g^2G|\phi|^2}}
  \bar A_\pm|\phi|\Biggr|^2       \nonumber \\
& &\phantom{\int d^2x  [}
 +{(1-2g^2|\phi|^2G)\over 4g^2|\phi|^2}
   \left(B\pm 2e g^2|\phi|^2{\cal F}({\cal F}^\prime)^2\right)^2
   \Biggr] \nonumber \\
& &\mp e{\cal F}_\infty\Phi
\mp 2\pi n({\cal F}(0)-{\cal F}_\infty)
\end{eqnarray}
where $n$ is the number of zeros and we have here chosen the potential $V$ as
\begin{equation}
V= e^2 g^2{\cal F}^2({\cal F}^\prime)^2.
\end{equation}
It is bounded from below as
\begin{equation}
E\ge | e{\cal F}_\infty\Phi|+ 2\pi |n({\cal F}(0)-{\cal F}_\infty)|,
\end{equation}
and the inequality is saturated by
a static configuration
satisfying the Bogomol'nyi equations
\begin{mathletters}
\label{bogom:all}
\begin{eqnarray}
B\pm 2e g^2|\phi|^2{\cal F}({\cal F}^\prime)^2=0, \label{bogom:a} \\
\partial_\pm|\phi|-i{e\over \sqrt{1-2g^2|\phi|^2G}}
\bar A_\pm|\phi|=0. \label{bogom:b}
\end{eqnarray}
\end{mathletters}
One can study various cases, corresponding to specific functional forms for
$G(|\phi|^2)$.
For the sake of simplicity, Ghosh \cite{gosh} choose $G$ as
\begin{equation}
G={1 -c_0(1-2g^2|\phi|^2)^{1-\gamma} \over 2g^2|\phi|^2}
\end{equation}
with two parameters, $c_0>0$ and a positive integer
$\gamma$. Then ${\cal F}$ becomes
\begin{equation}
{\cal F}=-{1\over \sqrt{c_0} g^2 (1+\gamma)}(1-2g^2|\phi|^2)^{(\gamma+1)/2}
\end{equation}
and the concurrent potential is
\begin{equation}
V=\left({e \over g c_0(1+\gamma)}\right)|\phi|^2 (1-2g^2|\phi|^2)^{2\gamma}.
\end{equation}
This potential has two degenerate vacuua,  the unbroken vacuum ( $|\phi|=0$)
and the broken vacuum ($|\phi|=1/\sqrt{2g^2}$).
In the unbroken vacuum,  ${\cal F}(0)$ and ${\cal F}_\infty$
have the same value, $-1/\sqrt{c_0} g^2 (1+\gamma)$,
and it leads to the lower bound for the energy functional given as
\begin{equation}
 {1 \over \sqrt{c_0} g^2(1+\gamma)}|e\Phi|.
\end{equation}
In the broken vacuum,  ${\cal F}(0)$ has the value
$-1/\sqrt{c_0} g^2 (1+\gamma)$ while
${\cal F}_\infty$ vanishes, thus the lower bound of the energy functional has
a quantized value
\begin{equation}
 {2\pi \over \sqrt{c_0} g^2(1+\gamma)}|n|.
\end{equation}
Detailed anaysis of other physical quantities can be found
in ref. \cite{gosh}.

We have derived the Bogomol'nyi equations for the solitons
in the generalized Maxwell Chern-Simons gauge theories with the magnetic
moment interaction as well as the usual minimal gauge interaction,
without assuming rotational symmetry a priori. In many cases the Bogomol'nyi
eqautions are related to the $N=2$ extended
supersymmetry \cite{super}. Along these lines we investigated some
$N=2$ supersymmetric models containing the magnetic momoment interaction,
but we could not find any direct relation between them and the models
studied here.

\acknowledgments

We would like to thank Professor Choonkyu Lee for
useful discussions.  This work was
supported in part by Korea Science and Engineering Foundation
(through  the Center for Theoretical Physics at Seoul National University),
and in part by the Basic Science Research Institute Program,
Ministry of Education, Korea (BSRI-94-2401 for TL and BSRI-94-2441 for HM).

\end{document}